# Intraseasonal characterization of tropospheric $O_3$ in the North of the Buenos Aires Province: determining four months cycle and teleconnection evidence.

Rodolfo G. Cionco[1], Rubén Rodriguez[1], Nancy Quaranta[1,2], Eduardo Agosta[3]

1- UTN-Facultad Regional San Nicolás
2- CIC Prov. Bs. As
3- Pontificia Universidad Católica de Argentina (UCA)-CONICET

**ABSTRACT**

Tropospheric ozone (O3T) is a secondary pollutant whose formation involved primarily solar radiation, NOx and volatile organic compounds. The North of the Buenos Aires Province (Argentina), has great agricultural-industrial activity; therefore, O3T study is an important issue in the area. In this paper, we present the first results tend to estimate and characterize O3T in San Nicolás de los Arroyos, North of Buenos Aires. Due to a lack of in situ data, we analyse the observations of the instrument OMI (Ozone Monitoring Instrument) of land remote sensing satellite AURA (GSFC/NASA). The data cover the years 2004-2013. Applying the multitaper technique (MTM), very suitable for short and noisy data series, spectral analysis is performed on a grid corresponding 1° in latitude by 1.5° in longitude, centred South of the Province of Santa Fe. The most remarkable result is the emergence of a significant peak (95%) of four months cycle. To test the validity of this signal in San Nicolás, daily solar radiation data ($Q$) were analysed in the area. The application of MTM to the daily values of $Q$, yields a spectral peak of 120 days. It is concluded that atmospheric opacity on the site has four months variations that modify the solar radiation at the troposphere and consequently, the production rate of O3T. Evidence that these variations are due to teleconnection process originated in the Maritime Continent are presented.

# Caracterización intraestacional de O3 troposférico en el norte de la Prov. de Bs. As.: determinación de un ciclo cuatrimestral

Rodolfo G. Cionco[1], Rubén Rodriguez[1], Nancy Quaranta[1,2], Eduardo Agosta[3]

1- UTN-Facultad Regional San Nicolás
2- CIC Prov. Bs. As
2- Pontificia Universidad Católica de Argentina (UCA)-CONICET

**RESUMEN**
El ozono troposférico ($O_3$T) es un contaminante secundario en cuya formación intervienen principalmente la radiación solar, NOx y compuestos orgánicos volátiles. El norte de la Provincia de Buenos Aires (Argentina) presenta gran actividad agrícola-industrial por lo tanto, el estudio de $O_3$T es un tema importante en la zona. En este trabajo se presentan los primeros resultados tendientes a estimar y caracterizar $O_3T$ en San Nicolás de los Arroyos, norte de Buenos Aires. Careciéndose de datos obtenidos in situ, se analizan las observaciones del instrumento OMI (Ozone Monitoring Instrument) del satélite de teleobservación terrestre AURA (GSFC/NASA). Los datos cubren los años 2004-2013. Aplicando la técnica de multitaper (MTM), muy adecuada para series cortas y ruidosas, se realiza el análisis espectral correspondiente sobre una grilla de 1° en latitud por 1,5° en longitud, centrada en el sur de la Provincia de Santa Fe. El resultado más notable es la aparición de un pico significativo (al 95 %) de periodicidad cuatrimestral. Para contrastar la validez de esta señal en San Nicolás, se analizaron datos de radiación solar diaria $Q$ obtenidos en la zona. La aplicación de MTM a los valores diarios de $Q$, arroja un pico espectral de 120 días (con menor potencia). Se concluye que la opacidad atmosférica del lugar presenta variaciones cuatrimestrales que modifican la radiación solar a nivel de la troposfera y en consecuencia la tasa de producción de $O_3T$. Se discuten diferentes mecanismos relacionados con el origen de este fenómeno.

## MOTIVACIÓN

En San Nicolás de los Arroyos, norte de la Provincia de Buenos Aires (Fig. 1), se realiza una intensa labor en monitoreo de calidad de aire por parte del Grupo de Estudios Ambientales de la UTN, con fuerte énfasis en contaminación vehicular e industrial. Además de las fuentes de origen humano, la contaminación del aire está sujeta a factores no locales dependientes (en diferentes escalas) de la meteorología y la climatología de la zona [1] y de procesos de teleconexión [2].

El $O_3$ superficial y troposférico en general, es también un tema importante en el lugar. Careciéndose de datos obtenidos in situ, se analizan las observaciones del instrumento OMI del satélite de teleobservación AURA. Los datos son mensuales y cubren los años 2004-2013 (Fig. 2). La producción de $O_3$ en la atmósfera depende fuertemente de la radiación solar, por lo tanto un ciclo anual es esperado. Sin embargo, variaciones de más corto período son potencialmente importantes por el posible impacto medioambiental. Períodos de 20 días han sido reportados como onda estacionaria o viajera para $O_3$ estratosférico [3] y para el contenido total de $O_3$ [4]. Sin embargo, no hay reportes indicando variaciones intraestacionales más específicas para $O_3T$.

## RESULTADOS

La (Fig.2) muestra en forma evidente las variaciones estacionales con los picos máximos indicados. La recta de regresión no indica una tendencia creciente significativa. Para evaluar la aparición de oscilaciones intraestacionales, se realiza un análisis espectral múltiple a la serie $O_3T$. Primero, se aplica multitaper (MTM), [5]) a la serie cruda (Fig. 3). El resultado (además de la gran banda anual) muestra una oscilación de ~ 4 meses, significativa al 95%. Aplicando máxima entropía (MEM, Fig. 3) se obtienen picos similares (3,9-4,3 meses). Disponiéndose de datos de irradiancia solar (sólo 3 años, 2006-2008), se analizaron también valores de radiación solar *diaria* (*Q,* Fig. 4). Notablemente, MTM muestra una periodicidad cuatrimestral (120 d), pero significativa al 90% y marginal al 95% (Fig. 5), junto a otro pico mayor de 100

d. MTM es una técnica muy adecuada para series cortas y ruidosas; MEM es muy precisa para hallar "líneas" en espectros. Como test para evaluar si esos picos de 4 meses son reales o debido al ruido inherente a la señal, se aplicó descomposición SSA (singular spectrum analysis) para eliminar ruido rojo y analizar el espectro resultante con MEM nuevamente [5]. El resultado (Fig. 6) muestra una banda muy significativa, cercana a 4 meses, aun con 10 polos (la serie $Q_3T$ es de 100 datos). La reconstrucción SSA (componentes 5-6 con períodos 3,8-4,6 meses) se muestra en la Fig.7. Particularmente se observa una buena representación de la serie para los años 2007 y mediados de 2008 (orden 28-50), a pesar del escaso número de eigenvectors utilizados.

## DISCUSIÓN Y CONCLUSIONES

El análisis anterior muestra que el pico de 4 meses es "real" con gran probabilidad. La aparición de la misma periodicidad en las series de $Q$ sugiere que el $O_3T$ está influenciado por la oscilación de $Q$ (esto es, que la radiación solar a nivel troposfera en la zona, modula la producción de $O_3T$). La cuestión es: ¿qué modula a $Q$? Mientras existen varios procesos que afectan al $O_3T$ en determinado sitio: quema de biomasa, producción de $NO_2$, ingresos estratosféricos; la radiación solar se ve influenciada por procesos convectivos a gran escala.

Nuestra hipótesis es que dentro de la escala intraestacional observada, sería posible determinar la existencia de propagación de ondas cuasi-estacionarias de Rossby, generadas por convección anómala tropical desde la región del Índico-Pacífico hacia el sur de Sudamérica. Estas oscilaciones de escala planetaria parecen modificar a $Q$ debido a procesos locales de nubosidad y/o precipitación, tal como se deduce de la Fig.8. Se concluye que la presencia de $O_3T$ en la zona norpampeana podría estar siendo modulada a escala intraestacional por procesos de teleconexión provenientes del Continente Marítimo.

## REFERENCIAS

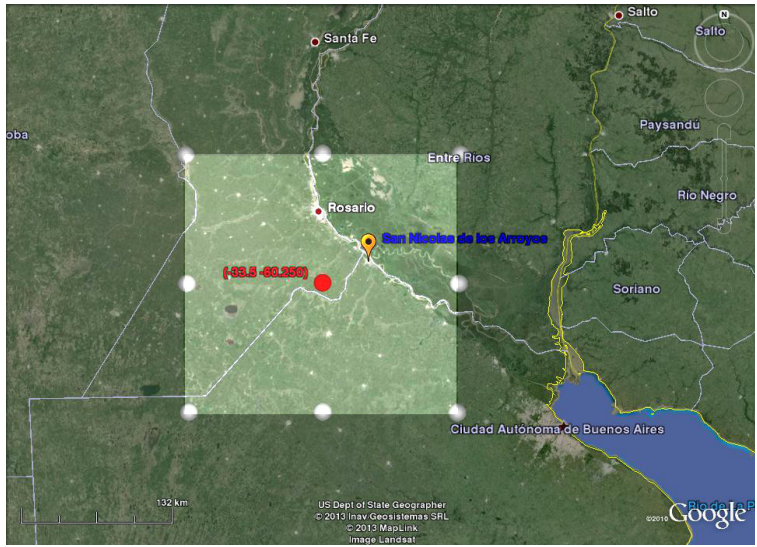

***Fig. 1***. *Mapa indicando la zona estudiada y la localidad bonaerense de San Nicolás de los Arroyos, en la República Argentina.*

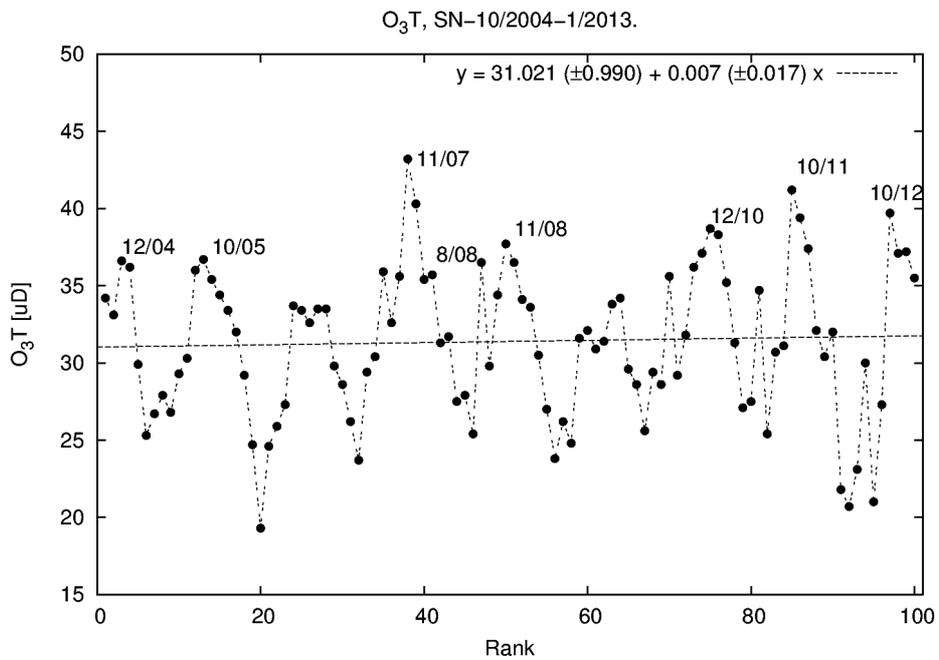

***Fig. 2***. *Serie de O3T estudiada. UD: unidades Dobson.*

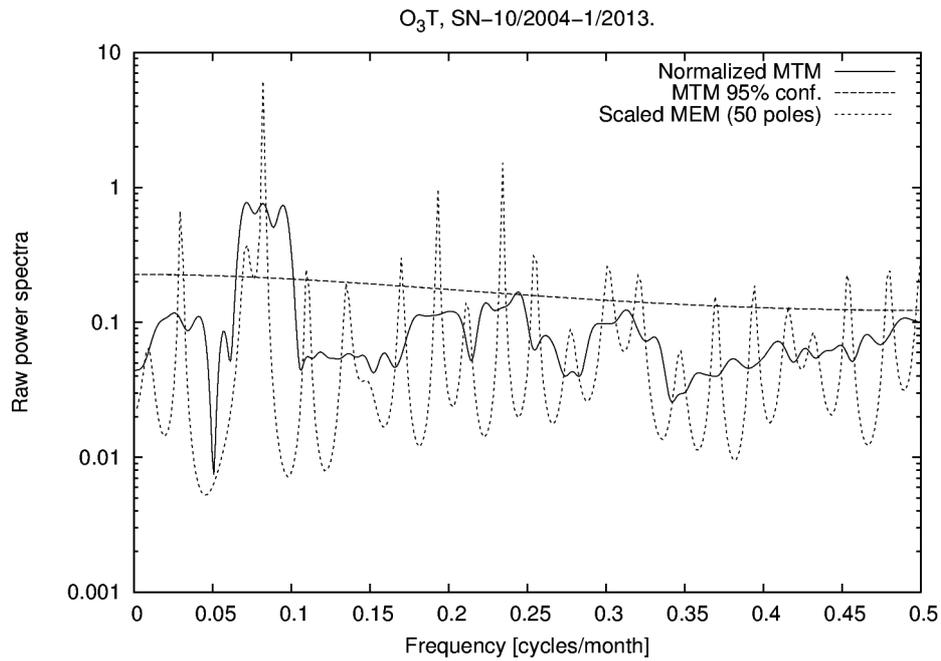

**Fig. 3**. Espectro de MEM y MTM de la serie $O_3T$ indicando potencia significativa alrededor de 4 meses.

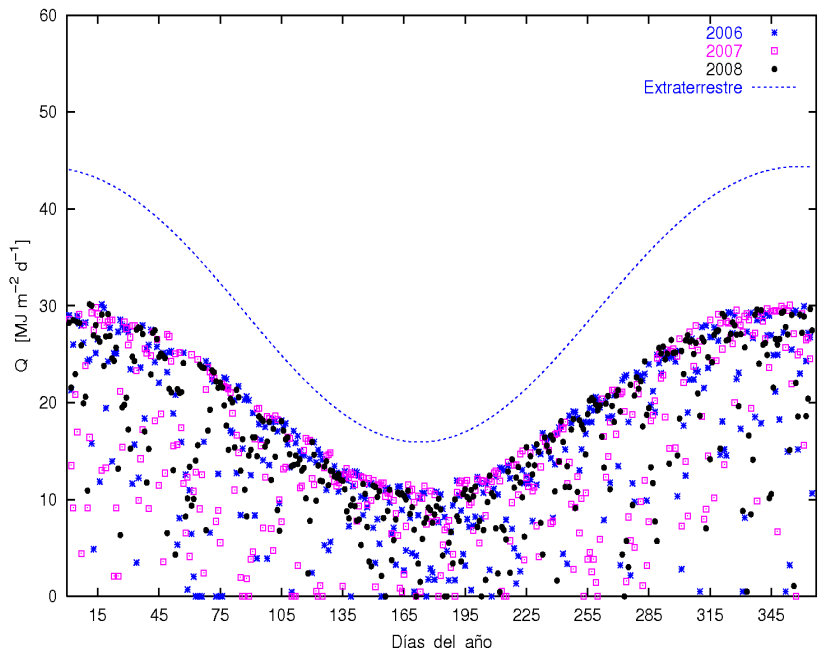

**Fig. 4**. Radiación solar diaria (Q) medida para los años indicados. Se muestra también el valor teórico por fuera de la atmósfera (línea punteada).

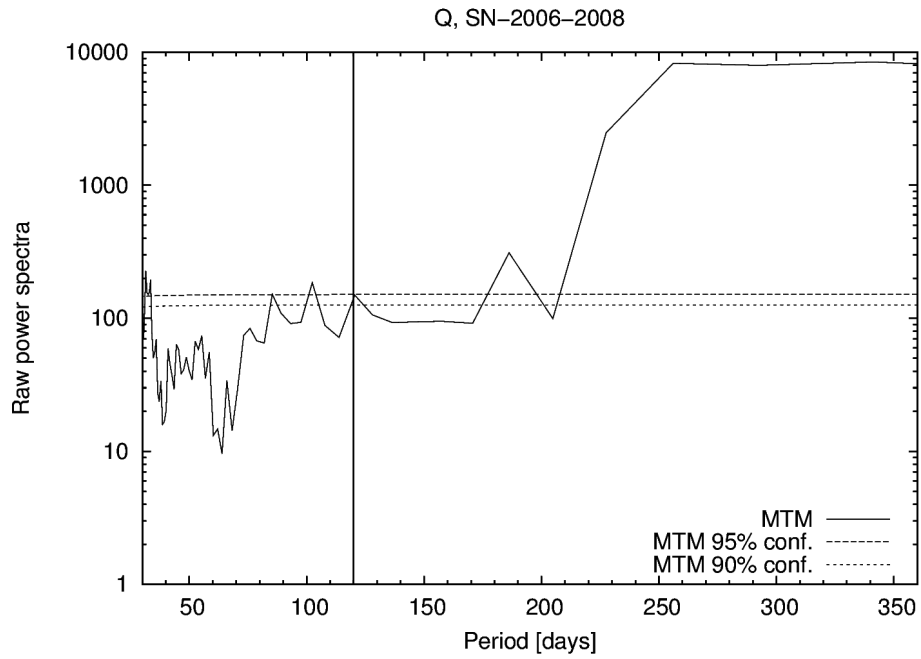

*Fig 5. Espectro MTM de la serie de Q solar. La línea vertical indica periodicidad de 4 meses.*

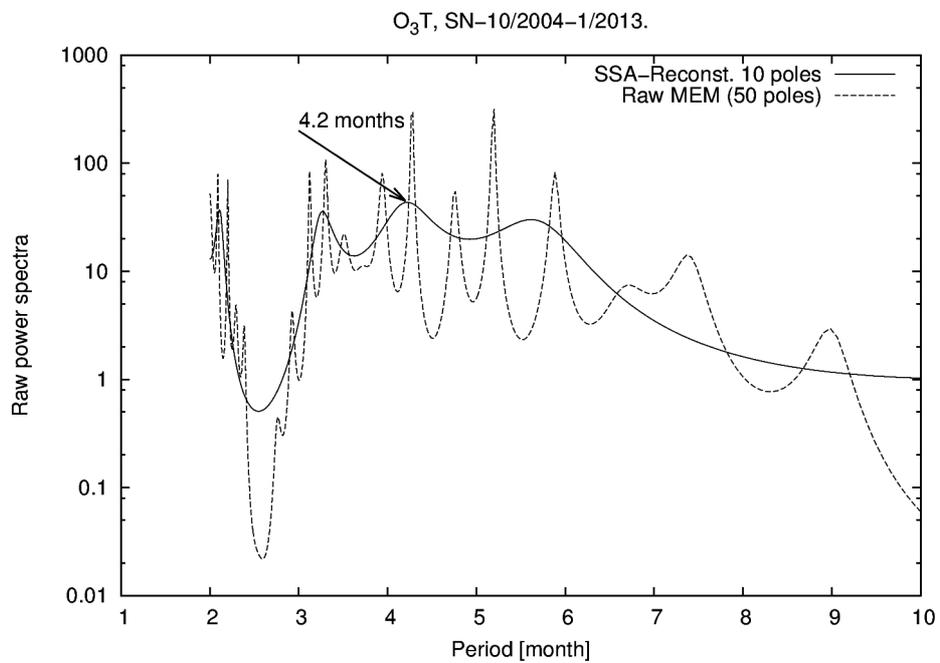

*Fig. 6. Espectro MEM de la serie O3T reconstruida mediante SSA. Una banda alrededor de 4.2 meses aparece con M = 10 = número de datos/10.*

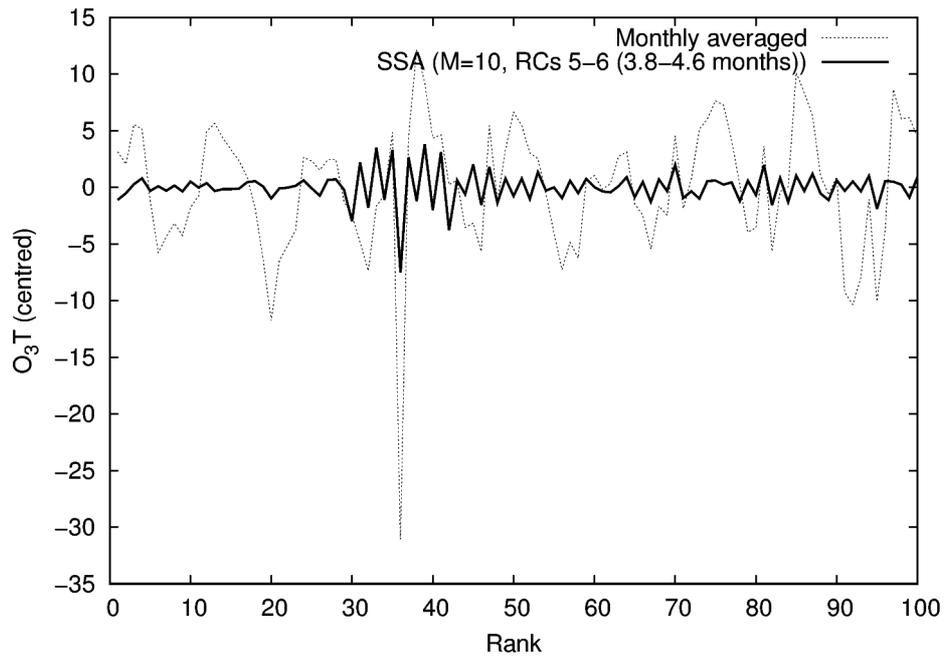

*Fig. 7. Serie de $O_3T$ reconstruida por SSA (componentes principales 5-6 con períodos cercanos a 4 meses).*

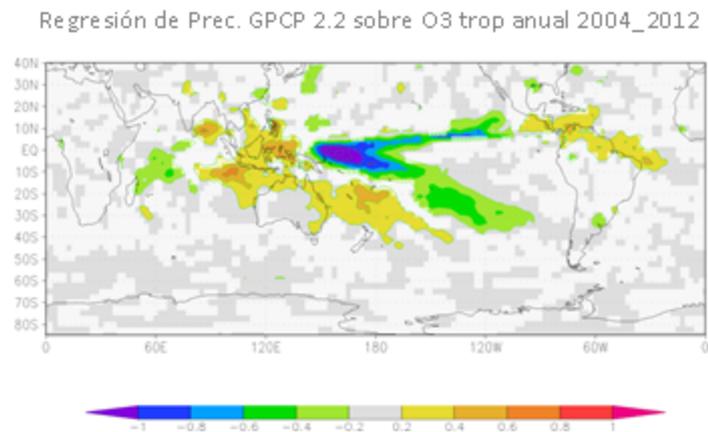

*Fig. 8. Análisis de regresión de precipitación sobre la serie O3T (promedios anuales).*